\documentclass[twocolumn]{aastex62}
\shorttitle{PRODUCTION OF EUV LATE PHASE FLARES}
\shortauthors{Dai \& Ding}

\begin{document}

\title{Probing the Production of Extreme-ultraviolet Late-Phase Solar Flares by Using the Model Enthalpy-Based Thermal Evolution of 
Loops} 

\author[0000-0001-9856-2770]{Yu~Dai}
\affiliation{School of Astronomy and Space Science, Nanjing University, Nanjing 210023, China}
\affiliation{Key Laboratory of Modern Astronomy and Astrophysics (Nanjing University), 
Ministry of Education, Nanjing 210023, China}
\affiliation{Key Laboratory of Space Weather, National Center for Space Weather, China Meteorological Administration, 
Beijing 100081, China}

\author{ Mingde~Ding}
\affiliation{School of Astronomy and Space Science, Nanjing University, Nanjing 210023, China}
\affiliation{Key Laboratory of Modern Astronomy and Astrophysics (Nanjing University), 
Ministry of Education, Nanjing 210023, China}

\correspondingauthor{Yu~Dai}
\email{ydai@nju.edu.cn}

\begin{abstract}
Recent observations in extreme-ultraviolet (EUV) wavelengths reveal an EUV late phase in some solar flares that is characterized by
a second peak in warm coronal emissions ($\sim3$~MK) several tens of minutes to a few hours after the soft X-ray (SXR) peak. 
Using the model enthalpy-based thermal evolution of loops (EBTEL), in this paper we numerically probe the production of EUV 
late-phase solar flares. Starting from two main mechanisms of producing the EUV late phase, i.e., long-lasting cooling and secondary heating,
we carry out two groups of numerical experiments to study the effects of these two processes on the emission characteristics in late-phase 
loops. In either of the two processes an EUV late-phase solar flare that conforms to the observational criteria can be numerically 
synthesized. However, the underlying hydrodynamic and thermodynamic evolutions in late-phase loops are different between the two
synthetic flare cases. The late-phase peak due to a long-lasting cooling process always occurs during the radiative cooling phase, while 
that powered by a secondary heating is more likely to take place in the conductive cooling phase. We then propose a new method for 
diagnosing the two mechanisms based on the shape of EUV late-phase light curves. Moreover, from the partition of energy input, we discuss
why most solar flares are not EUV late flares. Finally, by addressing some other factors that may potentially affect the loop emissions, we also
discuss why the EUV late phase is mainly observed in warm coronal emissions.
\end{abstract}


\section{INTRODUCTION}
It is widely accepted that solar flares are a result of the rapid release of magnetic energy stored in the solar corona. Through the magnetic 
reconnection process \citep{Parker63}, the free magnetic energy is rapidly converted into plasma heating in the flare loops, acceleration of 
charged particles, and in some cases, bulk plasma motion, which is later on manifested as a coronal mass ejection (CME) propagating into the 
interplanetary space.

When the impulsive flare heating is terminated, the flare loops will naturally cool down. This cooling process, however, is far from a static.
 Powered by an intense flare heating, the flare loop can initially  be heated to a temperature as high as several tens of megakelvin. 
This builds up a huge temperature gradient along the loop, resulting in a large heat flux conducting down into the transition region 
(TR) and chromosphere during the early cooling stage. In addition to the bombardment by a nonthermal electron beam, this excess heat flux 
must also drive an upward flow from the TR/chromosphere \citep{Neupert68}, which is conventionally termed chromospheric evaporation 
\citep{Antiochos78}, otherwise the heat flux would be too high to be effectively radiated away in the TR/chromosphere \citep{Antiochos76}. 
As the loop temperature decreases and the density increases, radiation gradually becomes dominant. In a loop under radiative cooling, 
regions of lower temperatures cool faster than those of higher temperatures \citep{Antiochos80}, as expected from the shape of the 
radiative loss function. This difference in cooling rate weakens the pressure gradient along the loop that balances against gravity, which, 
except for a static catastrophic cooling in the very late stage \citep{Cargill13}, consequently leads to a downward flow into the
TR/chromosphere \citep{Bradshaw05}, usually known as chromospheric condensation.  

Observationally, the evolution of  temperature and density in flare loops can be quantitatively diagnosed via the electromagnetic emissions 
from the loops in a wide wavelength range. According to the standard two-ribbon solar flare model, which is often called the CSHKP model 
\citep{Carmichael64,Sturrock66,Hirayama74,Kopp76}, the evolution of a solar flare can be divided into two phases: an impulsive phase, and 
a following gradual phase. The impulsive phase is characterized by a rapid increase of the emissions in hard X-ray (HXR) and 
chromospheric lines (e.g., \ion{He}{2}), indicating a prompt response of the solar lower atmosphere to the nonthermal electron 
bombardment and/or thermal conduction caused by the initial heating. Hot evaporated material then fills the flare loops, which brighten up
in soft X-ray (SXR) and coronal lines. In particular, the loop emissions peak in extreme-ultraviolet (EUV) wavelengths in the sequence of 
decreasing temperatures, constituting the gradual phase \citep{Chamberlin12}. 

In general, during the gradual phase of a solar flare, the EUV emission exhibits just one main peak shortly after the \emph{GOES} SXR 
peak. Nevertheless, by using full-disk integrated EUV irradiance observations with the EUV Variability Experiment \citep[EVE;][]{Woods12}
on board the \emph{Solar Dynamics Observatory} \citep[\emph{SDO};][]{Pesnell12}, \citet{Woods11} discovered an ``EUV late phase" in 
some flares, which is seen as a second peak in the warm coronal emissions (e.g., \ion{Fe}{16}, $\sim3$~MK) several tens of minutes to a
few hours after the \emph{GOES} SXR peak. There are, however, no significant enhancements of the SXR or hot coronal emissions 
(e.g., \ion{Fe}{20} and higher, $>10$~MK) in the EUV late phase, and spatially resolved imaging observations such as from the 
Atmospheric Imaging Assembly \citep[AIA;][]{Lemen12} on board the \emph{SDO} reveal that the secondary late-phase emission comes from a 
set of higher and longer loops that are anchored in the same flare-hosting active region (AR) as the original flaring loops. 

In a statistics of 191 solar flares higher than the \emph{GOES} C2 class that occurred during the first year of \emph{SDO} normal operations, 
\citet{Woods11} found that only 25 of them (13\%) exhibited an EUV late phase, and more interestingly, about half of the EUV late-phase flares 
took place in a cluster of two ARs, implying a specific magnetic configuration of the ARs in which EUV late-phase flares are preferentially
produced. Case studies  showed that EUV late-phase flares occur in  a multipolar magnetic field, which exhibits either a  classic or 
asymmetric quadrupolar configuration \citep{Hock12,LiuK13}, or a parasitic polarity embedded in a large-scale bipolar magnetic field  
\citep{Dai13,SunX13,Masson17}. Such a magnetic configuration observationally facilitates the existence of two sets of loops that are distinct in 
length rather than loops with a continuous length distribution, as further confirmed by force-free coronal magnetic field extrapolations
 \citep{Jiang13,SunX13,LiYD14,GuoY17,Masson17}. 

The discovery of the EUV late phase imposed a great challenge to current empirical flare-irradiance models
\citep[e.g.,][]{Tobiska00,Chamberlin08}, since they use the emission measured in SXRs as a flare proxy, and therefore would improperly
estimate the total flare energy if there exists an EUV late phase in the flare. However, currently, only a few reports on the EUV late phase are 
available in the literature, and the origin of this phase is still not fully understood.  Based on the observational facts mentioned above, some 
authors suggested that the EUV late phase might be due to a secondary energy injection into the long late-phase loops that is considerably 
delayed from the main flare heating \citep{Woods11,Hock12,Dai13}, while others proposed that both the main flaring loops and late-phase loops
are heated nearly simultaneously during the main flare heating, and the delayed occurrence of the EUV late phase is the result of a long-lasting
cooling process in the  long-late phase loops \citep{LiuK13,Masson17}. It was also pointed out that in some EUV late-phase flares the two 
mechanisms may both play a role \citep{SunX13}. 

Through numerical experiments using the model called enthalpy-based thermal evolution of loops (EBTEL), which was first proposed by 
\citet{Klimchuk08} and later improved by \citet{Cargill12} and \citet{Barnes16}, \citet{LiYD14} numerically studied the role of the above two
mechanisms in producing an EUV late phase, finding that a long cooling process in the late-phase loops can well explain the presence of 
the EUV late-phase emission, although the possibility of an additional heating in the decay phase cannot be excluded. Nevertheless, until 
now the EUV late-phase emission characteristics under different mechanisms have not been studied in depth to date, but deserve further
investigation. As a time-efficient model, the EBTEL model is very suitable for such studies. Using the EBTEL model,  we here 
explore the physical relationship between the emission characteristics and the underlying hydrodynamics and thermodynamics in different 
flare loops, and try to  answer the important questions why EUV late-phase flares only occupy a small fraction in all solar flares and why the
EUV late phase is mainly observed in warm coronal emissions. The paper is organized as follows. In Section 2 we briefly describe  the 
EBTEL model. In Section 3 we carry out two groups of numerical experiments, and use the outputs to synthesize different EUV late-phase 
flares. The numerical results are discussed in detail in Section 4, and a brief summary is presented in Section 5. 
 
\section{The EBTEL MODEL}
The EBTEL model is a zero-dimensional (0D) hydrodynamic model that describes the evolution of the \emph{average} temperature, 
pressure, and density along a coronal loop. When compared with those far more sophisticated one-dimensional (1D) hydrodynamic 
numerical simulations, EBTEL gives quite good agreement in average parameters, but is costs far less computation time. 

The basic idea behind EBTEL is that the imbalance between the heat flux conducting down into the TR and the radiative loss rate there 
leads to an enthalpy flux at the coronal base: an excess heat flux drives an evaporative upflow, whereas a deficient heat flux is 
compensated for by a condensation downflow.  This upward/downward enthalpy flow controls the hydrodynamic evolution of the 
coronal loop. We note that this idea has also been adopted to treat the ``unresolved TR"  in the recent 1D modeling in \citet{Johnston17}. 

If the flows along the loop are subsonic (which should hold for the most of the time of the evolution), the kinetic energy can be neglected and the 
equation of momentum can be dropped from the governing equations. By integrating the equations of continuity and energy conservation
over the coronal section of the loop, which has a half-length $L$ measured from the coronal base to apex (EBTEL assumes a symmetric 
semi-circular loop geometry), we obtain 
\begin{equation}
\frac{dn}{dt}=-\frac{c_2}{5c_3kTL}(F_0+c_1\mathcal{R}_c)
\end{equation}
and
\begin{equation}
\frac{dp}{dt}=\frac{2}{3}\left[Q-(1+c_1)\frac{\mathcal{R}_c}{L}\right],
\end{equation}
where $n$, $p$, and $T$ are the average density, pressure, and temperature of the coronal loop, respectively, $k$ is the Boltzmann 
constant, $c_2$ ($c_3$) is the ratio of the average coronal (coronal base) temperature to apex temperature, 
$F_0=-(2/7)\kappa_0(T/c_2)^{7/2}/L$ (where $\kappa_0=8.12\times10^{-7}$ in cgs units is the classical Spitzer thermal conduction 
coefficient) is the heat flux at the coronal base, $\mathcal{R}_c={n}^2\Lambda(T)L$ (where $\Lambda(T)$ is the optically thin radiative
loss function) approximates the radiative loss rate from the corona, $c_1$ is the ratio of radiative loss rate of the TR to that of the corona, 
and $Q$ is the average volumetric heating rate. For brevity, we omit the word ``average": so if not explicitly specified, hereafter quantities 
like density, pressure, temperature, and heating rate refer to the loop-averaged quantities. The evolution of the coronal temperature then 
follows from the equation of state, which is 
\begin{equation}
\frac{1}{T}\frac{dT}{dt}=\frac{1}{p}\frac{dp}{dt}-\frac{1}{n}\frac{dn}{dt}.
\end{equation}

Of the three parameters $c_1$, $c_2$, and $c_3$ in EBTEL,  $c_1$ is the most important. Analytical calculations based on 
the static coronal loop model in \citet{Martens10} gave a value for $c_1$ of around 2, and values for $c_2$ and $c_3$ close to 0.9 and 0.6, 
respectively. In the original EBTEL model \citep{Klimchuk08}, to achieve an overall consistency with the 1D simulation results, $c_1$ 
was kept as a constant of 4 throughout the loop evolution. Nevertheless, when a loop evolves dynamically,  the equilibrium is broken
and $c_1$ will also evolve. Obviously, the choice of a fixed $c_1$ value is not physically reasonable. In the later EBTEL versions, more 
physics was included to more appropriately determine the values of $c_1$.  One piece of the physics previously absent is gravitational
stratification, whose main effect is to depress the coronal radiation. Thus, higher values of $c_1$ can be expected, especially for long 
loops with significant ratios of the length to the gravitational scale height \citep{Cargill12}. Another one piece is the deviation of the 
dynamically cooling loop from equilibrium states. In the early conductive cooling phase, the density increase in response to the
strong heat flux is not so fast that the loop is under-dense with respect to a static loop at the same temperature, leading to higher
values of  $c_1$ \citep{Barnes16}. During the radiative cooling phase, the loop is instead overdense, which will in turn reduce
the $c_1$ values \citep{Cargill12}.  In our following numerical experiments, we consistently calculate $c_1$ by taking all 
these factors into account, and hold $c_2$ and $c_3$ at their typical values of 0.9 and 0.6, respectively.
 
\section{NUMERICAL EXPERIMENTS} 
We use EBTEL to trace the hydrodynamic response of coronal loops to an impulsive energy release. For a prescribed loop half-length
$L$ and volumetric heating rate function $Q(t)$, we obtain the temporal evolution of the temperature $T(t)$, density $n(t)$, and pressure $p(t)$ 
of a coronal loop by numerically solving the time-dependent EBTEL  Equations~(1)--(3). Then we use  the outputs to synthesize 
the EUV emissions of the loop as they are in real solar observations. The overall light curves of a solar flare can be synthetically generated 
by combining the emissions of a series of loops  with different lengths and/or energy injections. We have carried out two groups 
of numerical experiments to probe the effects of varying the prescribed parameters on the loop emission characteristics and their 
implications in the production of an EUV late-phase flare. 
 
\subsection{Experiment 1: Effect of the Loop Length}
Previous case studies of EUV late phase flares showed that the EUV late-phase emission originates from a second set of higher and longer 
loops \citep{Hock12,Dai13,LiuK13,LiuK15,SunX13,Masson17}. The long-lasting cooling process explanation suggests that the EUV late-phase 
loops are heated nearly simultaneously with the main flaring loops, and their delay in warm EUV emissions is attributed  to a longer
cooling process for greater loop lengths \citep{LiuK13,LiYD14,Masson17}.  In EBTEL numerical experiment 1, we therefore vary the loop 
half-length and check its influence on the emission properties of the loops. 

\begin{deluxetable*}{lccccccc}
\tablecaption{Parameters in the EBTEL Numerical Experiments}
\tablehead{\colhead{No.} & \colhead{$L$} & \colhead{$A_{\mathrm{pix}}$} & \colhead{$Q_0$} & \colhead{$\tau_H$} & \colhead{$Q_b$} 
& \colhead{$T(t=0)$} & \colhead{$n(t=0)$}\\
\colhead{}  & \colhead{($10^9$ cm)}  & \colhead{(AIA pixel)}  & \colhead{(erg cm$^{-3}$ s$^{-1}$)}  
&\colhead{(s)}  &\colhead{(erg cm$^{-3}$ s$^{-1}$)}  &\colhead{(MK)} &\colhead{(10$^8$ cm$^{-3}$)}}
\startdata
Experiment 1 & & & & & & & \\
$i=1$ (M1) & 1.5 & 1 & 1.500 & 300 & $10^{-5}$ &0.44 & 0.86\\
$i=2$ & 3.0 & 1 & 0.750 & 300 &$10^{-5}$ & 0.66 & 0.89 \\
$i=3$ & 4.5 & 1 & 0.500 & 300 &$10^{-5}$ & 0.83 & 0.90 \\
$i=4$ & 6.0 & 1 & 0.375 & 300 &$10^{-5}$ & 0.97 & 0.89 \\
$i=5$ (L1) & 7.5 & 1 & 0.300 & 300 &$10^{-5}$ & 1.11 & 0.89 \\
$i=6$ & 9.0 & 1 & 0.250 & 300 &$10^{-5}$ & 1.23 & 0.88 \\
$i=7$ & 10.5 & 1 & 0.214 & 300 &$10^{-5}$ & 1.34 & 0.87 \\
$i=8$ & 12.0 & 1 & 0.188 & 300 &$10^{-5}$ & 1.45 & 0.86 \\
$i=9$ & 13.5 & 1 & 0.167 & 300 &$10^{-5}$ & 1.56 & 0.86 \\
$i=10$ & 15.0 & 1 & 0.150 & 300 & $10^{-5}$ & 1.66 & 0.86 \\
Experiment 2 & & & & & & & \\
$i=1$ (L1) & $7.5$ & 1 & $0.300$ & 300 &$10^{-5}$ & 1.11 & 0.89\\
$i=2\cdots4$ & $7.5$ & $10^{(i-1)/2}$ & $3\times10^{-(i+1)/2}$ & 300 &$10^{-5}$ & 1.11 & 0.89\\
$i=5$ (L2) & $7.5$ & $100$ & $0.003$ & 300 &$10^{-5}$ & 1.11 & 0.89\\
\enddata
\tablecomments{The columns show the loop half-length (Column 2),  normalized loop cross-sectional area (Column 3), 
peak rate of the triangular heating pulse (Column 4), duration of the pulse (Column 5), background heating rate (Column 6), initial 
loop temperature (Column 7), and density (Column 8).}
\end{deluxetable*} 

The parameters in experiment 1 are listed in the upper rows of Table 1. Here we consider a total of 10 loops, whose half-lengths 
consecutively increase from $1.5\times10^9$~cm to $1.5\times10^{10}$~cm with an even step of $1.5\times10^9$~cm. Note that the 
specification of $L$ values is not arbitrary but  based on observations. Loops lying toward the short end of our loop cases have lengths 
typical of main flaring loops, whereas those considerably longer (length ratio $>$ 2) correspond to late-phase 
loops \citep{Dai13,LiuK13,SunX13}. The initial equilibria in the loops are maintained by a steady background heating with a heating rate 
$Q_b$ of $10^{-5}$~erg~cm$^{-3}$~s$^{-1}$. As shown in the last two columns of Table 1, the initial loop density is almost a constant 
close to $9\times10^{7}$~cm$^{-3}$ in the different loops, while the initial temperature increases monotonically from 0.44 MK to 
1.66 MK as the loop half-length increases. This temperature increase tendency can be mathematically expected from the general 
equilibrium solutions of EBTEL, which reveal a scaling relationship as $T\sim L^{4/7}$. Generally speaking, if a short loop were initially 
heated to the same  temperature as in a longer static loop, a too large temperature gradient would be built up along the short 
loop, with the resulting excess heat flux in turn quickly smoothing this sharp temperature gradient.

Starting from the initial equilibria, we impose an impulsive heating on the loops, whose heating rate $Q$ has a  triangular profile with a 
total duration $\tau_H$ of 300 s (with equal ascending and descending durations), comparable to the  duration of the impulsive phase in a 
typical solar flare. The peak values  $Q_0$ of the  heating rate are set in such a way that for the $i$th loop ($i=1\cdots10$),
$Q_0=1.5/i$~erg~cm$^{-3}$~s$^{-1}$. This guarantees that the total (length-integrated) energy deposition rate $2QL$ is the same for 
all the loops, whose peak value is $4.5\times10^9$~erg~cm$^{-2}$~s$^{-1}$, also typical for a solar flare \citep{LiY12,LiYQ14,Qiu12}. 

\begin{figure}
\plotone{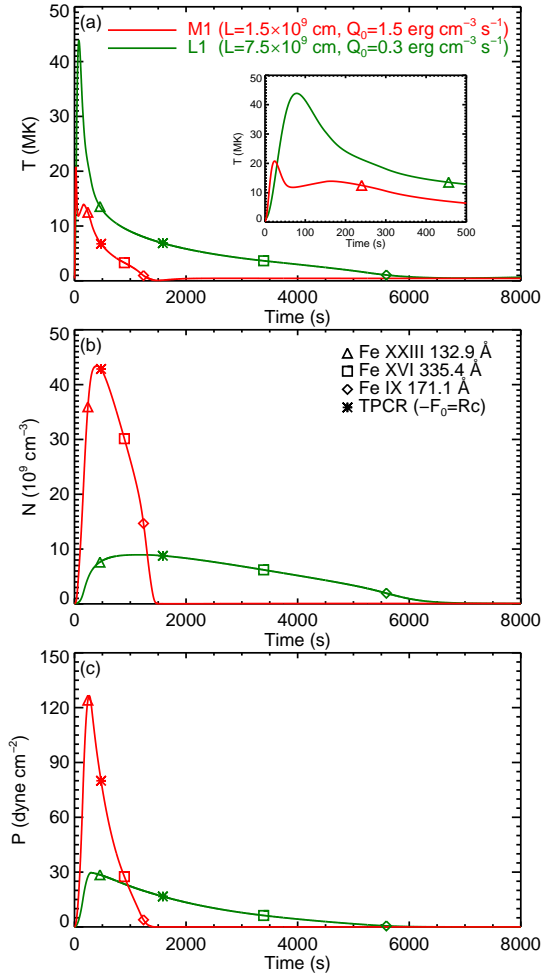}
\caption{Temporal profiles of the temperature (a), density (b), and pressure (c) in loops M1 (red) and L1 (dark green). The inset in panel (a) 
gives a zoom-in view of the temperature evolution during the first 500~s. The triangle, square, and diamond mark the properties at the 
times of line irradiance peaks in \ion{Fe}{23} 132.9~{\AA}, \ion{Fe}{16} 335.4~{\AA}, and \ion{Fe}{9} 171.1~{\AA}, respectively, and the 
asterisk outlines the transition point from conductive-dominated cooling to radiative-dominated cooling (TPCR) in loop.}
\end{figure}

Figure 1 shows the temporal evolution of the temperature, density, and pressure in two representative loops (the first loop in experiment 1, 
typically a main flaring loop, hereafter M1, and the fifth loop in the experiment, typically a late-phase loop, hereafter L1) in response to the 
corresponding heating pulse. It is clearly seen that the short loop M1 evolves much faster than the long loop L1. This evolution pattern 
holds when all other loops are included in the experiment. Of the three parameters, the temperature reaches its maximum first, and the
density peaks last. Interestingly, the pressure maximum occurs very close to the end time of the heating pulse at $t=300$~s. For the 
reason mentioned above, longer loops can attain a higher maximum temperature than shorter loops. As to the density and pressure, it is 
noted that the two peak values are roughly inversely proportional to the loop half-length.

Based on the output temperature and density, we then synthesize the emissions of the loops in several optically thin EUV lines and
passbands as they are observed  from a near-Earth perspective (e.g., the \emph{SDO}). The total loop irradiance in a specific line (as that
observed in EVE) is given by
\begin{equation}
I_{\mathrm{Line}}=0.83Ab(X)C(T)\langle \mathrm{EM}\rangle \frac{A}{4\pi d^2} \ 
(\mathrm{erg}\ \mathrm{cm}^{-2}\ \mathrm{s}^{-1}),
\end{equation}
where the coefficient 0.83 is the commonly assumed ratio of the hydrogen density to electron density in the solar corona, 
$Ab(X)$ is the abundance of element $X$ relative to hydrogen, $C(T)$ is the contribution function of the emitting ion under
consideration, which can be calculated using the CHIANTI atomic database \citep{DelZanna15}, $\langle\mathrm{EM}\rangle=2{n}^2L$ is 
the spatially averaged column emission measure (EM) of the loop, $A$ is the loop cross-sectional area, and $d$ is the Sun-Earth distance. 
Meanwhile, the total intensity of the loop in an AIA passband can be computed as 
\begin{equation}
I_{\mathrm{AIA}}= R(T) \langle \mathrm{EM}\rangle  A_{\mathrm{pix}}\ (\mathrm{DN}\ \mathrm{s}^{-1}),
\end{equation}
where $R(T)$ is the temperature response function of the AIA passband, and $A_{\mathrm{pix}}$ is the normalized loop cross-sectional 
area in units of AIA pixel. In this work, we consider three iron lines in EUV wavelengths: \ion{Fe}{23} 132.9~{\AA}, \ion{Fe}{16} 335.4~{\AA},
and \ion{Fe}{9} 171.1~{\AA}, whose peak formation temperatures are 14.1~MK,  2.8~MK, and 0.9~MK, reflecting emissions from hot flaring 
plasmas, warm corona, and cool background corona, respectively. For a comparison with the EUV line irradiance, we choose the AIA 
passbands of 131~{\AA}, 335~{\AA}, and 171~{\AA} (hereafter AIA 131, 335, and 171), respectively, the dominant ion of which is either the
same as or of a similar formation temperature to that of the corresponding EUV line \citep{ODwyer10}. 

The commonly used AIA response functions $R(T)$, like those distributed within the AIA package of Solar Software (SSW), are generated  
using a coronal abundance \citep[e.g.,][]{Fludra99} by default. Moreover, the radiative loss function $\Lambda(T)$ used in EBTEL is 
also formulated based on the coronal abundance. For consistency, when using Equation (4) to calculate irradiance of the three iron lines, we 
adopt  a typical abundance of iron in the solar corona, e.g., $7\times10^{-5}$. This will enable us to cross-check data from both EVE and AIA 
when modeling a real solar flare.

Previous observations have revealed that the effective width of a coronal loop is on the order of 1 Mm \citep[e.g.,][]{Aschwanden08}. With 
the improvement of spatial resolution, it is believed that finer structures within a loop can be resolved with future instruments. In this sense, the 
specification of a loop cross-sectional area is somewhat arbitrary. For convenience, in experiment 1 the cross-sectional area of all loops is set
 to one AIA pixel size, i.e., $0\farcs6\times0\farcs6$, which yields a nominal solid angle $A/d^2$ of $8.46\times10^{-12}$ sr in Equation~(4) and
a unity of $A_{\mathrm{pix}}$ in Equation~(5). In practice, this assumption has also been adopted in modeling thousands of flare loops of a 
solar flare in AIA observations \citep[e.g.,][]{Qiu12,Zhu18}, where the chromospheric light curve (e.g., AIA 1600 {\AA}) of each brightening pixel 
in flare ribbons is used to construct the heating function in the half-loop (of a cross-sectional area of also one AIA pixel) anchored on that pixel. 
Choosing other values of $A$ ($A_{\mathrm{pix}}$) does not affect our main results, since according to Equations (4) and (5), it just 
changes the amplitude of the synthetic light curves by the same factor. In passing, we note that under this assumption, the total energy of the 
impulsive heating injected into each loop in our experiment is $\sim1.29\times10^{27}$~erg for $d=1.5\times10^{13}$~cm.

\begin{figure*}
\epsscale{1}
\plotone{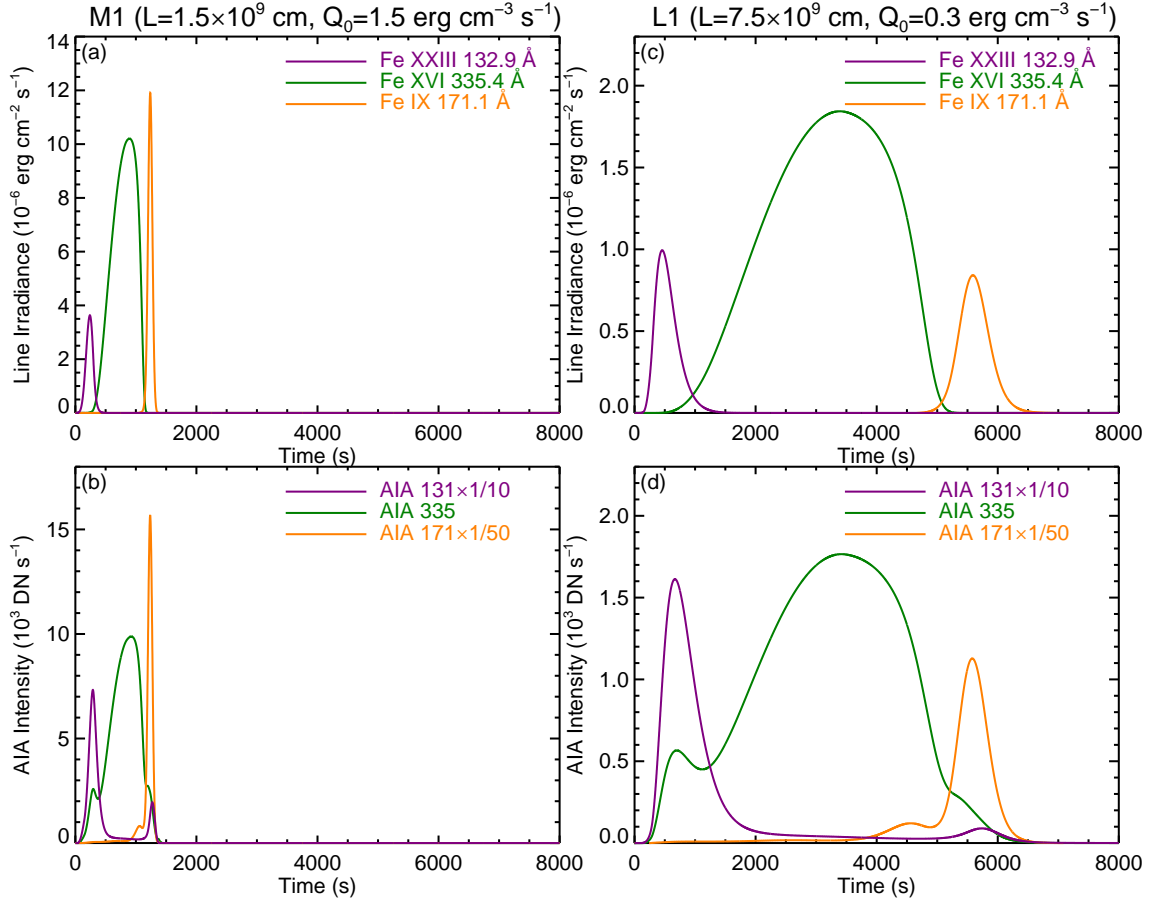}
\caption{Synthetic light curves of loops M1 (left) and L1 (right), with the background emission subtracted. The upper and lower panels show
the light curves in three EUV lines and three AIA passbands, respectively. The curves are color-coded and scaled according to the legend 
in each panel.}
\end{figure*}

Figure 2 displays the background-subtracted synthetic light curves of loops M1 and L1 in the three EUV lines and AIA passbands. 
For the individual loop, as the loop cools down, the loop emission peaks sequentially in lines/passbands of decreasing temperatures. 
When comparing the different loops, it is found that the increase in loop half-length not only delays the occurrence time of the
emission peak, but also reduces the amplitude of the peak. These effects are more prominent for the cool coronal emissions. 
For example, in loop M1, the largest irradiance peak occurs in the cool coronal \ion{Fe}{9} 171.1~{\AA} line, while in loop L1, the warm 
coronal \ion{Fe}{16} 335.4~{\AA} line contributes the most prominent irradiance, and the peak irradiance in \ion{Fe}{9} 171.1~{\AA} even 
drops to a level lower than that in the hot coronal \ion{Fe}{23} 132.9~{\AA} line. In general, the emission evolution in an AIA passband 
follows a consistent pattern to that in the corresponding EUV line. The peak times in AIA 335 and AIA 171 are nearly simultaneous (within 1
minute) with those in the  \ion{Fe}{16} 335.4~{\AA} and \ion{Fe}{9} 171.1~{\AA} lines. Nevertheless, the peak in AIA 131 shows a  
systematic delay up to 7 minutes with respect to the \ion{Fe}{23} 132.9~{\AA} peak, which we attribute to a slightly lower formation 
temperature of  \ion{Fe}{21} 128.8~{\AA} ($\sim$11.2 MK), the line of main contribution, compared to \ion{Fe}{23} 132.9~{\AA}, the line of 
much less contribution, in the hot band of AIA 131.

For a quantitative characterization of the emission pattern of the loops, we mainly focus on the line irradiance behavior because it more
physically reflects the underlying hydrodynamic evolution in the loops. We infer the peak times of the line irradiance from the two loops  
displayed in Figure 2. Meanwhile, we define the transition point from conductive-dominated cooling to radiative-dominated cooling (TPCR) 
for the loops, which can be mathematically expressed as the time when $-F_0=\mathcal{R}_c$ holds. The corresponding properties at 
these times are accordingly  overplotted on the  temporal profiles in Figure 1. In both loops, (1) the initial cooling 
process is dominated by conductive cooling,  which lasts until the loop temperature has considerably dropped but the density has just 
decreased slightly from its maximum; (2) the irradiance peaks in different lines lie in different cooling regimes: the peak in the hot 
\ion{Fe}{23} line occurs in the conductive cooling phase when the pressure is close to its maximum, while the peaks in the warm 
 \ion{Fe}{16} and cool \ion{Fe}{9} lines appear in the radiative cooling phase; and (3) the peaks in the \ion{Fe}{23} and 
\ion{Fe}{16} lines occur when the density is not far from its maximum, whereas the \ion{Fe}{9} line peaks  when the
material initially evaporated into the loop has been sufficiently drained.

\begin{figure*}
\epsscale{1}
\plotone{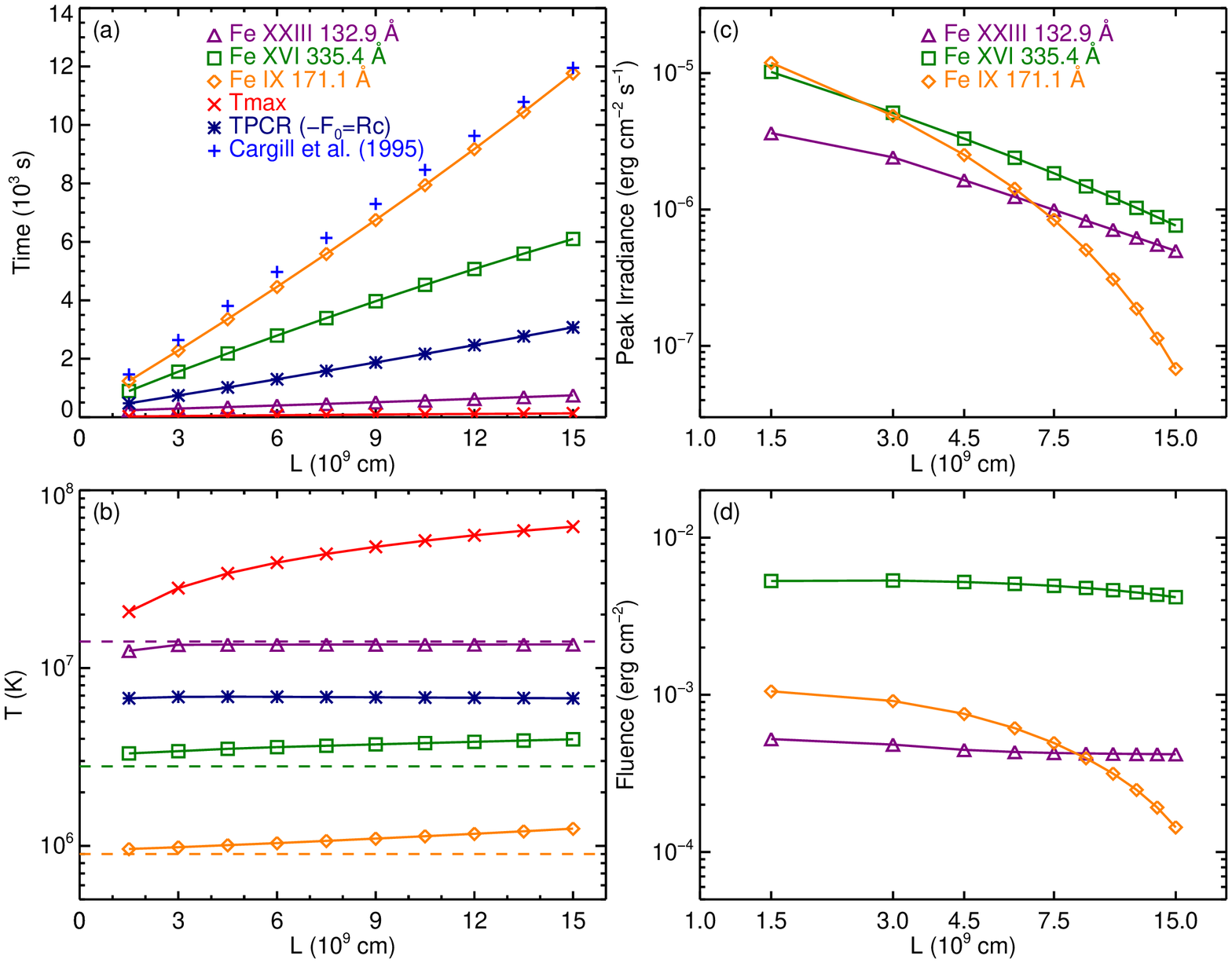}
\caption{Parameters characterizing the loop emission as functions of the loop half-length for the loops in EBTEL numerical experiment 1. 
Panel (a) shows the occurrence times of line irradiance peaks, loop maximum temperature, and TPCR, as well as the end time of loop 
cooling according to \citet{Cargill95}. Panel (b) shows the temperatures at the times of line irradiance peaks and TPCR, as well as the loop
maximum temperature itself.  The horizontal dashed lines mark the peak formation temperatures of the emitting 
ions. Panels (c) and (d) depict the background-subtracted peak irradiance and fluence of the three EUV lines, respectively. The meanings
of the symbols are explained in the legends.}
\end{figure*}

We further extend our study to all loops in experiment 1 whose emission characteristics as functions of the loop half-length are 
shown in Figure 3. As expected from the results based on the two-loop sub-sample, for all loops in the experiment, the irradiance 
peaks in hot and warm/cool lines are divided by the TPCR\@. The longer the loop, the later occurrence of the peaks, with the peak time
nearly linearly proportional to the loop half-length. Furthermore, the cooler the line, the more prominent the delay effect (Figure 3(a)).
By assuming a single power-law radiative loss function of $\Lambda(T)\sim T^{-1/2}$ \citep{Rosner78} and a temperature$-$density scaling 
law of $T\sim n^2$ during the mass-draining radiative cooling phase \citep{Serio91,Jakimiec92}, \citet{Cargill95} analytically evaluated the 
cooling time of a loop (finally cooling down to a temperature $T_L$ at which the single power-law of the radiative loss function becomes 
inaccurate), which is formulated in a ready-to-use form of $\tau_{\mathrm{cool}}=6.06\times10^{-5}{L^{5/6}}{{p_0}^{-1/6}}$~s,  where $p_0$ 
is the pressure at the start of the cooling. Here we substitute $p_0$ with $p_{\mathrm{max}}$, and rewrite the original formula as 
\begin{equation}
\tau_{\mathrm{cool}}=t_{p_{\mathrm{max}}}+6.06\times10^{-5}\frac{L}{(p_{\mathrm{max}}L)^{1/6}}\  (\mathrm{s}),
\end{equation}
where $t_{p_{\mathrm{max}}}$ is the time of maximum pressure, which is very close to $t=300$~s according to the numerical results in
experiment 1. The theoretically predicted end times of the cooling in our loop cases derived from Equation (6) are also plotted in Figure 
3(a), which show a close proximity to the \ion{Fe}{9} peak times. 

The temperatures at the times of line irradiance peaks are in general 
consistent with the characteristic formation temperatures of the emitting ions (Figure 3(b)). Nonetheless, systematic temperature deviations are 
also evident to some extent. In the conductive cooling domain, although the loops are all initially heated well above the peak formation 
temperature of \ion{Fe}{23}, the temperature at the time of \ion{Fe}{23} irradiance peak is systematically below it. In the radiative cooling 
regime, the temperatures at the times of \ion{Fe}{16} and \ion{Fe}{9} irradiance peaks are instead systematically higher than their peak 
formation temperatures, and furthermore, the deviations are increasingly stronger as the loop half-length increases. It is worth noting 
that in all loops radiation starts to dominate over the loop cooling at an almost constant temperature of around 6.8 MK\@. 

The background-subtracted peak irradiance for all three EUV lines drops monotonically as the loop half-length increases. (Note that the 
background levels are orders of magnitude lower than the peak values so that the effect of background-subtraction is actually negligible.) 
However, the irradiance decreases are quite different in the different lines, as reflected by the different slopes of the curves in 
Figure 3(c). Except for the first data point in each curve, the peak irradiance in both \ion{Fe}{23} and \ion{Fe}{16} very closely follows a 
power-law dependence  on the loop half-length (straight lines in the $\log-\log$ plotting). The fitted power-law index for \ion{Fe}{23} curve is 
$-0.98$, which means that the \ion{Fe}{23} peak irradiance is excellently inversely proportional to the loop half-length. The \ion{Fe}{16} 
curve is slightly more sloped, with a fitted index of $-1.18$. The \ion{Fe}{9} peak irradiance, however, deviates from the power-law 
dependence and decreases much more quickly, changing from the strongest  to the weakest irradiance with the increase in loop half-length. 

We also compute the line fluence by integrating the background-subtracted line irradiance over the whole loop evolution cycle. It is
found that the fluence of \ion{Fe}{23} and \ion{Fe}{16} remains at a relatively flattened level regardless of the loop half-length, while the 
\ion{Fe}{9} fluence drops by almost an order of magnitude (Figure 3(d)). Since the total amount of energy injected into each loop is 
constant, this means that the increase in loop half-length greatly reduces the contribution of the cool coronal emissions to the total radiation 
output from the loop while not obviously affecting those of the hot and warm coronal emissions. In the following,  these loop emission 
characteristics are further investigated in terms of an overall heating-cooling cycle of the loops.

\subsection{Experiment 2: Effect of the Heating Rate}
It has also been proposed that the EUV late phase is powered by a secondary energy release into the long late-phase loops, which
is considerably delayed from the main phase heating and therefore should be much less energetic \citep{Woods11,Hock12,Dai13}. In 
numerical experiment 2, we study the long loops. By adjusting the amplitude of the heating pulse, we probe how the heating rate
affects the loop emission properties.

The parameters in experiment 2 are listed in the lower rows of Table 1. The first loop in experiment 2 is exactly the same as loop L1 in
experiment 1, which we take as the reference loop. By fixing the loop half-length to be $7.5\times10^9$~cm, we consecutively reduce the 
peak rate of the impulsive heating by a factor of  3.16 ($10^{1/2}$). Meanwhile, to compensate for the decrease in the total energy 
deposition rate ($2QL$), we accordingly increase the loop cross-sectional area by the same factor. This factor is equivalent to the
number of loops/strands if we assume that the elementary loop has a constant cross-section, as mentioned above. Therefore, the total 
energy input is unchanged for all the cases in experiment 2, which is also the same as in experiment 1. Note that for the fifth and last 
loop in the experiment, which we refer to as loop L2, the amplitude of the heating pulse is still 300 times larger than the background heating 
rate.

\begin{figure*}
\epsscale{1}
\plotone{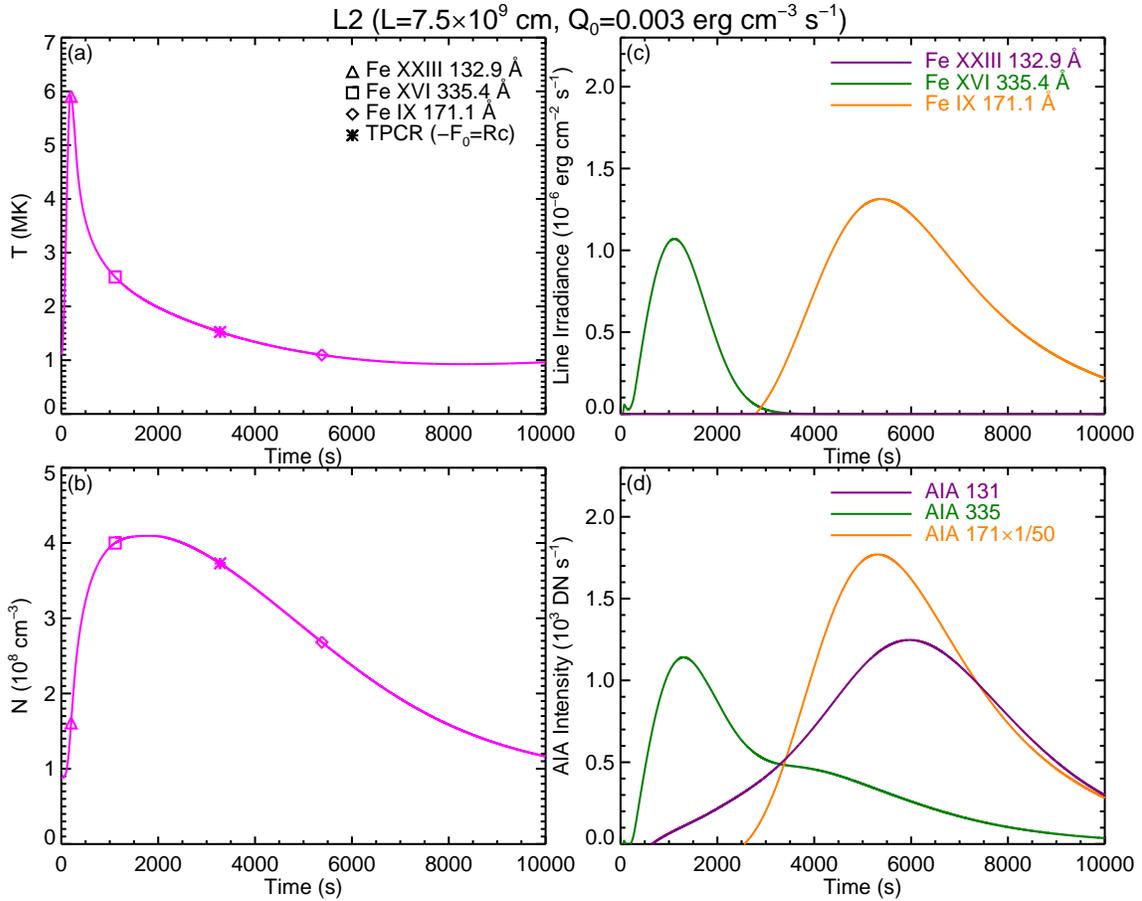}
\caption{Left: Temporal profiles of the temperature (a) and density (b) in loop L2. Right: Background-subtracted synthetic light curves 
of loop L2 in the three EUV lines (c) and AIA passbands (d). The left column is organized similarly to Figure 1 and the right column to
Figure 2.}
\end{figure*}

Figure 4 shows the temporal evolution of the temperature and density in loop L2, as well as the synthesized background-subtracted light 
curves in the three EUV lines and AIA passbands. Compared with loop L1 (Figure 1), the heating rate in loop L2 is 100 times lower, and 
therefore both the temperature and density there evolve relatively more slowly. More importantly, the amplitudes of increase in both 
quantities are much smaller in loop L2 (Figures 4(a) and (b)) than those in loop L1. The maximum temperature of loop L2 is only 5.9~MK, 
well below the typical temperatures of hot flaring plasmas ($>10$ MK).  

From the synthetic light curves, it is found that the hot \ion{Fe}{23} line irradiance from loop L2 is totally ``invisible" (although its peak 
irradiance is orders of magnitude larger than the  background level), as expected from the significantly lower loop temperature relative to 
the peak formation temperature of \ion{Fe}{23}\@. Regarding the emissions of lower temperatures, the peak irradiance of loop L2 in the 
warm  \ion{Fe}{16} line decreases slightly as compared with the case of loop L1, whereas the peak irradiance in the cool  \ion{Fe}{9} line 
increases slightly. As a result, the relative strengths of the two lines are just reversed for the two loops (Figure 4(c)). Interestingly, it is also
found that in loop L2, the peak time of \ion{Fe}{16} line irradiance moves closer to the time of impulsive heating, 2280~s earlier than that for 
loop L1, while the occurrence time of \ion{Fe}{9} irradiance peak remains relatively unchanged (within 300~s for the two loops). When we
overplot the corresponding properties at the times of line irradiance peaks and TPCR on the temporal profiles in Figures 4(a) and
(b), it is clearly seen that in loop L2, the \ion{Fe}{16} irradiance peak has moved to the conductive cooling regime, near the time of 
maximum density.  Although the \ion{Fe}{9} irradiance peak still remains in the radiative cooling phase, it is shifted somewhat toward the 
density maximum. 

Like that in experiment 1, the light curve in an AIA passband is generally consistent with that in the corresponding EUV line (Figure 4(d)). 
An exception occurs in AIA 131, in which the intensity peak even lags the AIA 171 peak, reflecting  a major contribution from the even cooler
lines \ion{Fe}{8} 130.9/131.2~{\AA} ($\sim0.6$~MK) rather than \ion{Fe}{21} 128.8~{\AA} in this passband under the condition of lower loop 
temperatures. Note that for the convenience of comparison, the light curves of loops L1 (Figures 2(c) and (d)) and L2 (Figures 4(c) and (d))
are plotted within the same ranges.

\begin{figure*}
\epsscale{1}
\plotone{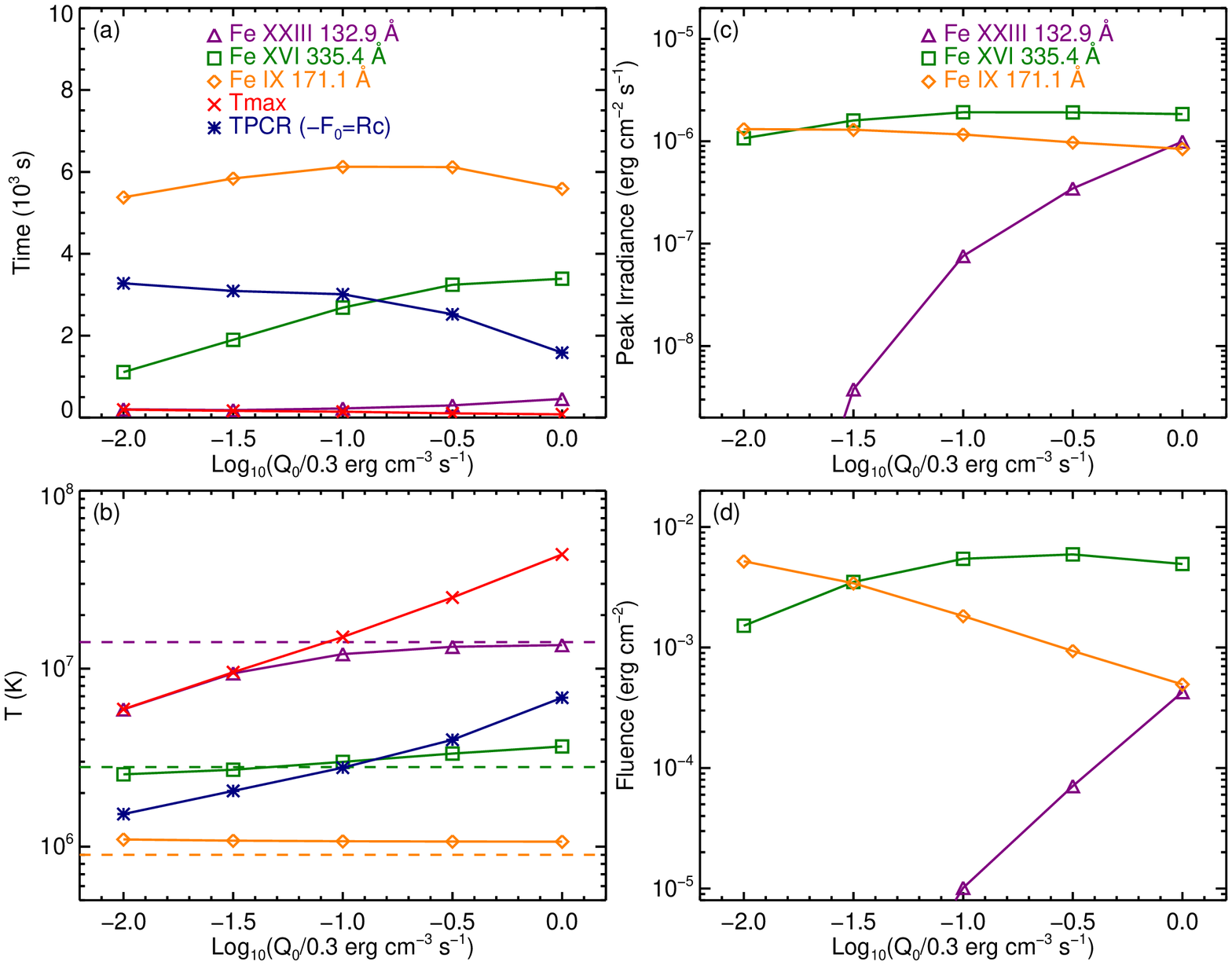}
\caption{Parameters characterizing the loop emission as functions of the peak heating rate for the loops in EBTEL numerical experiment 2. 
The panels are organized in the same way as in Figure 3.}
\end{figure*}

The emission characteristics as functions of the peak heating rate for the loops in experiment 2 are displayed in Figure 5. Note that the
rightmost data point  in each curve refers to the corresponding property in the reference loop L1. As the peak heating rate decreases, the 
peak time of \ion{Fe}{16} line irradiance moves monotonically toward the time of heating pulse, while the time of TPCR exhibits an 
opposite tendency. As a result, the location of the \ion{Fe}{16} irradiance peak experiences a transition from the radiative cooling domain to the
conductive cooling domain (Figure 5(a)). Following this transition, the temperature at the \ion{Fe}{16} irradiance peak decreases
accordingly, changing from slightly above the peak formation temperature of the line to slightly below it (Figure 5(b)). Since the loop 
maximum temperature drops dramatically with the decrease in heating rate, the occurrence of the \ion{Fe}{23} irradiance peak, if it could be 
observed, finally lies within the time period of the loop maximum temperature. The \ion{Fe}{9} irradiance peak, nevertheless, occurs at a
nearly constant temperature of $1.1$~MK, slightly but systematically higher than the peak formation temperature of the line; its 
occurrence time varies just in a narrow range in the radiative cooling regime.  

As to the amplitude of the emissions, the decrease in heating rate only marginally affects the peak irradiance of \ion{Fe}{16} and 
\ion{Fe}{9}\@. The peak irradiance of \ion{Fe}{16} first is constant and then decreases slightly, while that of \ion{Fe}{9} continues to
increase, although the increase extent is very small (Figure 5(c)). When considering the line fluence, this effect becomes obviously
exaggerated. As the peak heating rate decreases, the fluence of \ion{Fe}{16} drops by a factor of 3.3, whereas the \ion{Fe}{9} fluence 
increases by an order of magnitude (Figure 5(d)). In addition, with the decrease in heating rate, not only the peak irradiance, but also the 
fluence of \ion{Fe}{23} are depressed well below the detectable level.

\subsection{Synthesis of an EUV Late-Phase Flare}
The actual emission of a solar flare should consist of the contributions from a series of flare loops that have different lengths and/or 
undergo different energy release processes. Observationally, the length of a flare loop can be measured from spatially revolved 
imaging observations, and the heating function in that loop can be inferred from the chromospheric light curve at its footpoint, as done in 
\citet{LiYQ14} and \citet{Zhu18}. As a specific type of solar flares, previous case studies have pointed out that the emission 
of an EUV late-phase flare comes from two sets of flare loops distinct in length \citep{Hock12,Dai13,LiuK13,LiuK15,SunX13,Masson17}. 
For simplicity, in this work we just pick up one short main flaring loop and one long late-phase loop from our numerical experiments, and 
combine their emissions together to synthesize an EUV late-phase flare. Here our main concern is the general shape rather than the 
absolute amplitude of the flare light curves. 

Motivated by the two scenarios in producing the EUV late phase, we consider two cases. 
In the first case (hereafter Case 1), we directly add the background-subtracted synthetic light curves of loops M1 (Figures 2(a) and (b)) 
and L1 (Figures 2(c) and (d)) together, assuming that both loops are heated simultaneously and the partition of energy input is 1:1 
between the two loops. In the second case (hereafter Case 2), we select loops M1 and L2 (Figures 4(c) and (d)) and assume a time delay
in the heating between them. We thus first shift the light curve of loop L2 to a later time by 2000~s. This time delay is based on the time
difference of the \ion{Fe}{16} irradiance peaks (2280~s) between loops L1 and L2 found in numerical experiment 2. In this case, the total 
amount of energy injected into the late-phase loop L2 is also the same as that into the main flaring loop M1, although the total energy 
deposition rate in loop L2 is much lower.

\begin{figure*}
\epsscale{1}
\plotone{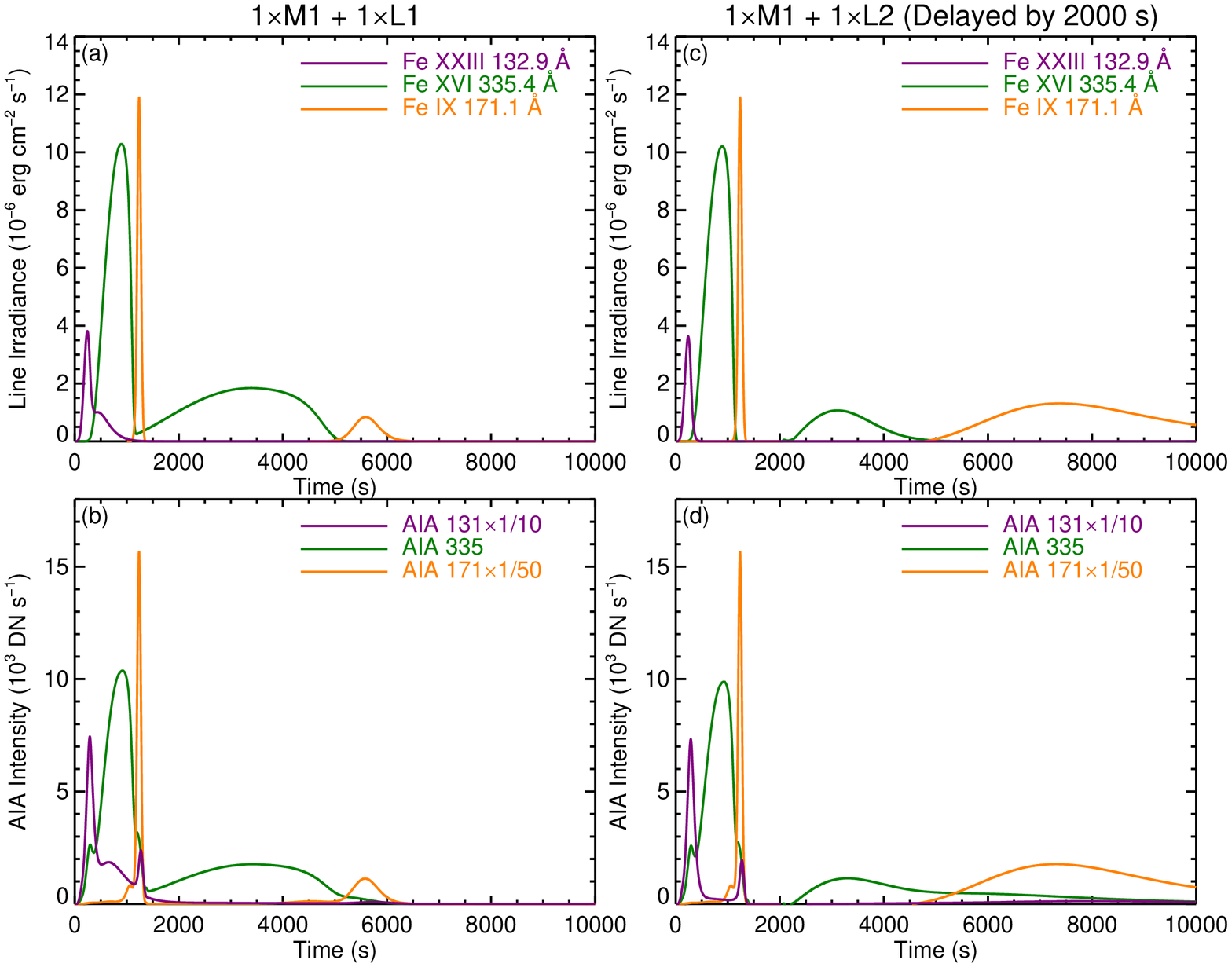}
\caption{Synthetic light curves of the flare loops mimicking EUV late-phase solar flares produced in two different processes. The light
curves are generated by adding the corresponding emission of loop L1 (Case 1, left) or L2 (Case 2, right) to that of loop M1, respectively. 
In Case 2, the light curve of loop L2 is assumed to be delayed by 2000~s. The upper and lower panels show the light curves in the three EUV lines and AIA passbands, respectively.  In all the curves, the background emission has been subtracted.}
\end{figure*}

The synthetic light curves of the two flare cases are shown in Figure 6. For the EUV line irradiance (Figures 6(a) and (c)), both cases 
exhibit an evident late phase-peak in the warm \ion{Fe}{16} line, which occurs 42 (37) minutes after the corresponding main flare peak with
an irradiance peak ratio of  0.18 (0.10) in Case 1 (2). These properties are in good agreement with those previously found both in statistics
\citep{Woods11} and case studies \citep{Hock12,Dai13,LiuK13,LiuK15,SunX13}. In spite of general similarity, there is also an obvious
difference between the \ion{Fe}{16} late phases in the two cases. In Case 1, the evolution of the late phase in \ion{Fe}{16} reveals a slow 
rise and then a fast decay, while the late-phase evolution in Case 2 shows an opposite pattern.  

It is also noted that both cases show a secondary late-phase peak in the cool \ion{Fe}{9} line that occurs 37 (71) minutes after the late 
phase peak of  \ion{Fe}{16} in Case 1 (2). Compared with that in Case 1, the late-phase of \ion{Fe}{9} in Case 2 is more prominent and 
prolonged. The irradiance in the hot \ion{Fe}{23} line, however, exhibits just one main flare peak in both cases. In Case 1, the \ion{Fe}{23} 
peaks in the short loop M1 and long loop L1 occur very close in time, therefore being merged into one main peak followed by a small bump
in the composite light curve. In Case 2, the \ion{Fe}{23} main peak is purely contributed by loop M1 because at that time the irradiance from
loop L2 is still at the background level. The late-phase heating, which goes into loop L2 more than half an hour later, is nevertheless not 
strong enough to power a detectable enhancement of the \ion{Fe}{23} irradiance again. 

As to the AIA intensities (Figures 6(b) and (d)), in both cases, the light curves generally resemble those in the corresponding EUV line. The
main difference is that there are two peaks in AIA 131 separated by an interval of only 16 minutes (in both cases). At  first glance, the 
second peak in AIA 131 may be regarded as a late-phase peak in the hot coronal emissions by mistake. A further investigation indicates that
this peak mainly comes from loop M1 and takes place very shortly after the main flare peak in AIA 171 (also see Figure 2(b)). Therefore, the
appearance of the two peaks in AIA 131 in fact reflects a broad coverage of both high and low temperatures in this passband 
\citep{ODwyer10}, as pointed out above.

\section{DISCUSSION}
Thanks to the high-quality observations in EUV wavelengths with both full spectral coverage and multiple passband imaging provided by
the recently launched \emph{SDO} mission, the EUV late-phase emission has been discovered, which is seen as a second peak in the 
warm coronal emissions well after the main flare peak \citep{Woods11}. These EUV emissions, combined with those formed at higher and 
lower temperatures, shed light on the hydrodynamic and thermodynamic evolution of the flare loops. The loop properties, such as 
temperature and density, are actually not uniformly distributed along a flare loop. Nevertheless, since we usually search for an EUV late 
phase from the loop-integrated light curves of a solar flare, the approach of adopting a 0D model to study the evolution of mean parameters 
of the flare loop is physically acceptable. There have been a number of 0D hydrodynamic models of coronal loops/strands 
\citep{Cargill12b}. In this work, we adopt the EBTEL model \citep{Klimchuk08,Cargill12,Barnes16} to numerically probe the production of 
EUV late-phase flares. Through a series of tests of the evolution of loops with different lengths and under different heating processes
\citep{Klimchuk08,Cargill12}, it has been proven that the EBTEL model gives results that agree quite well
with those from much more sophisticated 1D simulations such as ARGOS \citep{Antiochos99} and
HYDRAD \citep{Bradshaw03,Bradshaw11}, while the computation time is saved by orders of magnitude. In practice, the EBTEL model has
also been used  both in case studies \citep{Hock12,SunX13} and parametric surveys \citep{LiYD14} to reveal the nature of  the EUV late
phase. 

\subsection{Emission Characteristics of the Loops}
Based on the two explanations of the EUV late phase, i.e., long-lasting cooling and secondary heating, we have carried out two groups of 
numerical experiments to study the effects of these two processes on the loop emission characteristics. Figure 7(a) displays the 
temperature$-$density phase plot in an absolute scale for the loops in experiment 1, which depicts an overall heating-cooling cycle of 
the loops (in the clockwise direction): a fast temperature increase during the impulsive heating, followed by a conductive cooling with 
plasma evaporation and then a radiative cooling with mass draining, and finally a recovery to the initial equilibrium by the background 
heating. Although a higher maximum temperature is attained in a longer loop, the greater loop length increases the conductive cooling time 
$\tau_c$, assuming that the pressure is constant during the conductive cooling phase \citep{Antiochos78}. During the
radiative cooling phase, the lower density in the longer loop also results in a longer radiative cooling time $\tau_r$ \citep{Antiochos80}. 
When these two factors are combined, a significant difference in the overall cooling rate is expected between loops with distinct lengths, 
which naturally explains the delayed occurrence of an EUV late-phase emission.  

\begin{figure*}
\epsscale{1}
\plotone{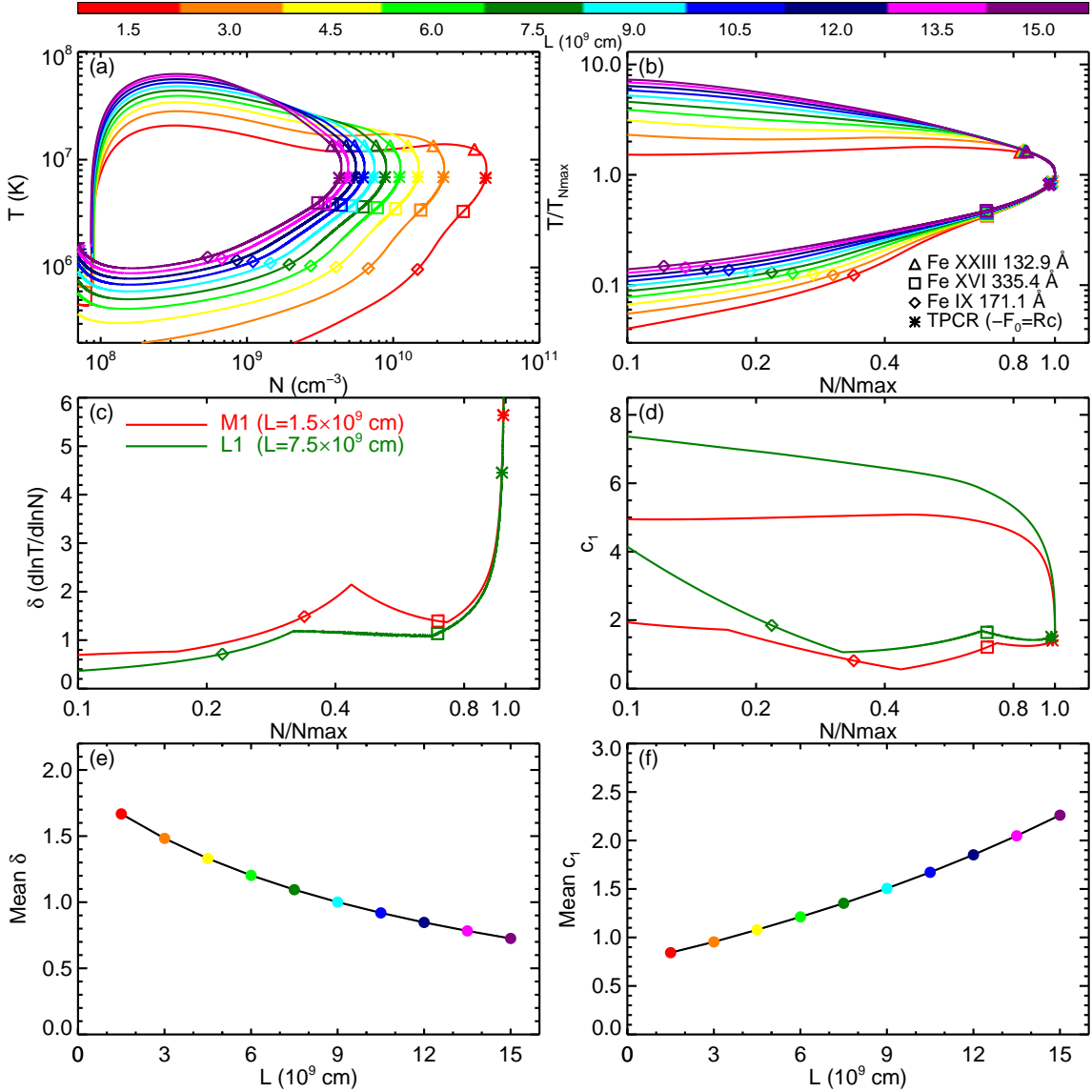}
\caption{Top: Temperature$-$density phase plot for the loops in EBTEL numerical experiment 1 in absolute scale (a) and normalized scale 
(b), respectively, where the temperature and density are normalized with respect to the corresponding values at the time of maximum 
density. Loops with different half-lengths are color-coded with the color bar attached above. The meanings of the symbols are 
explained in the legend of panel (b). Middle: Evolution of the power index $\delta$ (characterizing the scaling relationship 
$T\sim n^{\delta}$) during the radiative cooling phase (c) and the parameter $c_1$ in the EBTEL model (d) as functions 
of the normalized density for loops M1 (red) and L1 (dark green), respectively. Bottom: Average values of $\delta$ (e) and $c_1$ 
(f) for all loops over a time interval between the irradiance peaks of \ion{Fe}{16} (square) and \ion{Fe}{9} (diamond).}
\end{figure*}

According to Equation (4), the variation of line irradiance from a dynamically cooling loop is determined by the evolution of the 
temperature and density of the loop. In the conductive cooling phase, the density continues to increase when the temperature has 
passed through the value corresponding to the peak contribution function; thus the line irradiance will continue to increase until the 
role of the decrease in the contribution function finally overtakes the role of density increase. This causes the irradiance peak to appear 
at a temperature systematically below the peak formation temperature of the line, while in the radiative cooling phase, as the density 
continues to decrease, the line irradiance has already reached its peak before the loop cools down to the peak formation temperature
of the line. This emission pattern is consistent to what we have found in Figure 3(b).

Strictly speaking, according to Equation (1), the time when the density reaches its maximum should be defined as the time when
$-F_0=c_1\mathcal{R}_c$ is satisfied. This time is, indeed, very close to the time of TPCR (as can been seen in Figure 7(a)), 
since at this time, the parameter $c_1$ is close to 1 (as confirmed in Figure 7(d)). Quantitatively, the temperature at the time of 
maximum density lies in a narrow range around 8~MK for all loops in experiment 1, slightly higher than that (6.8~MK) at the time of 
TPCR\@. Therefore, it is reasonable to redraw the temperature$-$density phase plot by normalizing the density and temperature with 
respect to the corresponding values at the time of maximum density, as shown in Figure~7(b), in which all curves pass through a common point of $(1,1)$. 

As seen in the figure, all loops generally exhibit a self-similar evolution in the time period between the irradiance peaks of \ion{Fe}{23} and 
\ion{Fe}{16}\@. At the time of \ion{Fe}{23} irradiance peak, the column density of internal energy in the loops ($3pL$) is indeed very close
to the column density of total heating energy ($Q_0L\tau_H$), because the radiative loss is  almost negligible during that period (except 
for loop M1, where a considerable amount of energy has been radiated away). Considering that the temperature at the time of \ion{Fe}{23} 
irradiance peak is nearly the same among the loops, we can easily derive a density$-$length relationship of 
$nL\sim\mathrm{constant}$ and therefore a relationship of $\langle\mathrm{EM}\rangle=2{n}^2L\sim L^{-1}$ at this time. According
to Equation (4), this results in a perfect inverse relationship between the \ion{Fe}{23} peak irradiance and the loop half-length, as 
shown in Figure~3(c). The density$-$length relationship holds until the time of the \ion{Fe}{16} irradiance peak, at which the 
temperature becomes to differ slightly from each other and longer loops tend to have a higher temperature. In this situation, the 
difference in the contribution function also plays a role, causing the \ion{Fe}{16} curve in Figure 3(c) to slope slightly more. 

After the \ion{Fe}{16} irradiance peak, the loop evolutions begin to deviate from each other. For the same temperature decrease, the relative 
density decrease is more significant in longer loops, which can be inferred from the slopes of the curves in Figure~7(b). The 
density dependence of the curve slope $d\ln T/d\ln n$, which is equal to the power index $\delta$ of the scaling relationship 
$T\sim n^{\delta}$ during the radiative cooling phase, is shown in Figure~7(c) for loops M1 and L1. The values of $\delta$ for 
the long loop L1 are systematically lower than those for the short loop M1 during the whole radiative cooling phase. By averaging the 
values of $\delta$ over an interval between the times of corresponding peaks of \ion{Fe}{16} and \ion{Fe}{9}, we derive mean $\delta$ 
values of 1.67 and 1.09 for loops M1 and L1, respectively. Using the full 1D HYDRAD code, \citet{Bradshaw10} numerically studied the 
mass-draining radiative cooling process of loops with different lengths, revealing a $\delta$ value around 2 for short loops 
($2L=2\times10^9$~cm) and a reduced value to about 1 for very long loops (e.g., $2L=2\times10^{10}$~cm), which are in good agreement 
with our 0D EBTEL results. In addition, assuming that during the radiative cooling stage the TR radiation is purely maintained by a downward
enthalpy flow, \citet{Bradshaw10} also gave an analytical expression of $\delta$, which we rewrite as $\delta=\gamma-1+1/c_1$, where 
$\gamma=5/3$. This formula qualitatively  implies a larger $\delta$ caused by a smaller $c_1$, and vice versa. In Figure 7(d), we 
compare the evolution of the EBTEL parameter $c_1$ for loops M1 and L1, and find that this is indeed the case. During the whole loop 
evolution, the $c_1$ values for loop L1 are always higher than those for loop M1. The reason just lies in the new physics included in the 
improved EBTEL model. With a higher ratio of the loop length to the gravitational scale height, the radiative loss of loop L1 from the corona 
is more significantly depressed by the gravitational stratification, leading to higher values of $c_1$ in loop L1 than those in loop M1. 

In Figures 7(e) and (f), we plot the mean values of $\delta$ and $c_1$ for all loops in experiment 1, respectively. As expected, as the loop 
length increases, the mean $\delta$ value ($\bar{\delta}$) decreases monotonically from 1.67 to 0.73, while the mean $c_1$ value 
($\bar{c}_1$) increases monotonically from 0.84 to 2.26. We do not seek to build a quantitative relationship between $\bar{\delta}$ and 
$\bar{c}_1$, because the contribution of the heat flux has not been evaluated, although the role of thermal conduction in this stage should 
be marginal.

Since the loop evolution diverges after the \ion{Fe}{16} irradiance peak, the peak irradiance of \ion{Fe}{9} no longer follows a power-law 
dependence, but drops much more quickly with the increase in loop half-length (see Figure 3(c)), and the fluence of \ion{Fe}{9} is
also greatly depressed, as opposed that of to the \ion{Fe}{23} and \ion{Fe}{16} lines (see Figure 3(d)). For the same reason, the line irradiance 
in long loop peaks in the radiative cooling domain at a higher temperature than that in a short loop, as also shown in Figure 3(b).
  
In a case study of two EUV late-phase flares, \citet{LiuK13} used the \citet{Cargill95} formula to estimate the cooling times of the late-phase 
loops, which are qualitatively consistent with the observed time delays in \ion{Fe}{9} irradiance peak. In this work, we rewrite the 
original \citet{Cargill95} formula as Equation~(6). According to the parameters in experiment 1, at the start of the cooling (the time of
maximum pressure), the quantity $p_{\mathrm{max}}L$ is approximately constant among all the loops, except for loop M1 where 
$p_{\mathrm{max}}L$ is 15\% lower. This fluctuation of $p_{\mathrm{max}}L$ has only little influence on the cooling time. Therefore, a 
linear relationship between the cooling time and the loop half-length is expected, just as shown in Figure 3(a). Interestingly, the theoretically 
predicted end time of the loop cooling shows a close proximity to the time of \ion{Fe}{9} irradiance peak  for the corresponding experiment 
loop. In the EBTEL model, the radiative loss function is given in a piecewise continuous form \citep{Klimchuk08}, and the 
temperature$-$density scaling relationship changes from $T\sim n^2$ for short loops to $T\sim n$ for long loops according to our 
experiment. Both values are different from the assumptions adopted in \citet{Cargill95}. The surprisingly excellent consistency between the 
results of the two approaches implies that the shape of radiative loss function and gravity may have a significant effect on the amplitude of the 
loop parameters, but they do not considerably affect the timing of the loop evolution. Although EBTEL is just a simplified 0D model, it 
can still help us capture some essential characteristics of the flare evolution in case studies.

\begin{figure}
\plotone{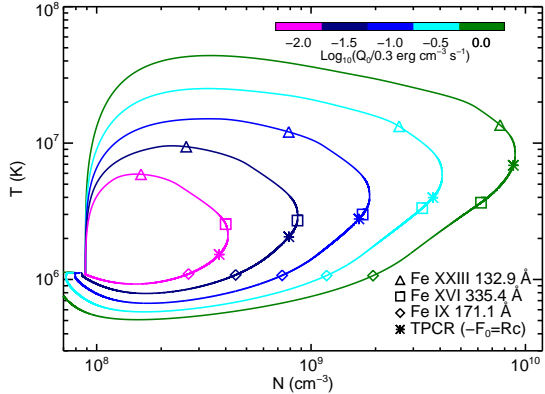}
\caption{Temperature$-$density phase plot for the loops in EBTEL numerical experiment 2 in absolute scale. Loops heated with different 
amplitudes are color-coded. The meanings of the symbols are the same as in Figure 7.}
\end{figure}

Figure 8 displays the temperature$-$density phase plot in an absolute scale for the loops in experiment 2.  The 
maximum temperature itself and the temperatures at the times of maximum density and TPCR both drop considerably as the heating rate 
decreases, as also shown in Figure 5(b). Therefore, normalization of the temperature$-$density phase plot, as done for the loops in 
experiment 1, is physically not applicable in this experiment. To be precise, quantitatively probing the emission characteristics of loops in 
experiment 2 is not as straightforward as what we have done for the loops in experiment 1.  Qualitatively, we propose that temperature
is a decisive factor in this experiment. 

As the loop maximum temperature decreases to well below those of hot flaring plasmas, i.e., $\sim10$~MK, the  irradiance of hot coronal
lines, e.g., \ion{Fe}{23}, is totally depressed to an undetectable level. Nevertheless, even with the least energetic heating, the loop can still
attain a sufficiently high maximum temperature of about $6$~MK, which guarantees the appearance of a sufficiently prominent 
irradiance peak in the warm \ion{Fe}{16} line. Different from that in experiment 1, the \ion{Fe}{16} irradiance peak
has gradually moved to the conductive cooling domain in experiment 2. Since in the conductive cooling phase the cooling rate is 
relatively faster, the peak time of \ion{Fe}{16} irradiance closer approaches the time of heating pulse, indicating a 
prompt response of the warm coronal emissions to the late-phase heating. Meanwhile, the temperature at  the time of \ion{Fe}{16} 
irradiance peak drops below the peak formation temperature of the line.  Considering that the radiation output from the loop has now moved
to lower temperatures, the irradiance of the cool \ion{Fe}{9} {line} is certainly enhanced. All these emission characteristics are shown in 
Figure 5.  

Finally, according to Equation (6), the \ion{Fe}{9} peak time should be postponed as the heating rate decreases, since the quantity 
$p_{\mathrm{max}}L$ decreases accordingly. However, this is not the case in experiment 2, as seen in Figure 5(a). The reason is that the 
condition for the \citet{Cargill95} formula to hold, $T_H\gg T_L$, (where $T_H$ and $T_L$ are the temperatures at the start and end of the
cooling, respectively,) is no longer satisfied.

\subsection{Production of the EUV Late Phase}
By combining the emissions from a short main flaring loop and a long late-phase loop in our numerical experiments, we have 
synthesized the light curves of two cases mimicking the EUV late phase flares produced in two different processes. As seen in 
Figure 6, both cases exhibit an EUV late-phase in the warm \ion{Fe}{16} line. Case 1 is based on the long-lasting cooling scenario. The 
difference in cooling time for \ion{Fe}{16} between the two loops of distinct lengths is indeed large enough to separate the late-phase 
peak from the main flare peak, while for the hot \ion{Fe}{23} line, the peaks in two loops are merged into one because the cooling rate is 
still very fast in the long loop during the conductive cooling phase. However, a small bump is produced following the main flare peak in the 
composite \ion{Fe}{23} light curve. Since the SXR emissions are formed at a similar temperature to that of \ion{Fe}{23}, we would see a
similar dual-decay behavior in the corresponding \emph{GOES} SXR light curve. We note that this dual-decay behavior in \emph{GOES} 
SXRs has been used as a proxy for EUV late-phase flares  prior to the\emph{SDO} era \citep{Woods14}. Case 2 is in accordance with the 
secondary heating scenario. The late-phase heating produces a rather prompt enhancement of the \ion{Fe}{16} emission in the long 
late-phase loop. However, the amplitude of the heating is too small to raise the temperature to that required for the formation of 
detectable \ion{Fe}{23} emission. 

As seen from the temperature$-$density phase plot in Figures 7 and 8, the density at the \ion{Fe}{16} irradiance peak in the late-phase 
loop is not far from its maximum value in both cases. The large enough density guarantees a sufficiently strong \ion{Fe}{16} late-phase
peak, which can be readily observed. In addition, we also note that the late-phase peaks in the two cases occur in different cooling 
domains. In Case 1, the late-phase peak occurs during the radiative cooling phase. After this, the irradiance is notably affected by the fast 
mass draining, therefore showing a fast decay. In Case 2, however, the late phase-peak takes place during the conductive cooling phase 
when the density is still increasing.  The decay phase in Case 2 is thus more gradual than the rise phase. \citet{LiYD14} have proposed two 
preliminary methods to determine whether or not a secondary heating plays a role in the late-phase emission. Here, we propose a new 
method to diagnose the processes behind an EUV late phase, which is based on the shape of light curves of warm coronal emissions. 
Light curves with a slow rise followed by a fast decay favor the long-lasting cooling scenario, whereas those showing an opposite pattern
are more likely due to the secondary heating process. 

We have applied this method to diagnosing a series of EUV late-phase flares 
occurring in 2011 September from NOAA AR 11283. The production of EUV late phase in the X2.1 class eruptive flare on  
September 6 has been investigated by \citet{Dai13}, who found convincing evidence for a late energy injection into the the late-phase loops. 
The late-phase light curves in both EVE 335 {\AA} and AIA 335 {\AA} (Figures 1 and 4 in their paper) reveal a more gradual decay 
than the rise. In another M1.2 class non-eruptive flare on 2011 September 9 (Y.Dai et al. 2018, in preparation), it is found the the late-phase
loops undergo an intense heating even 5 minutes earlier than the main flare heating. Both the overall late-phase light curve and the light curve 
extracted from a single late phase loop in \ion{Fe}{16} show a relatively faster decay than the rise. These evolution patterns are in
quality consistent with our expectation, and the detailed results will be presented in a separate paper.
 
Even with an equal energy partition between the main flaring loop and late-phase loop, the irradiance ratios of the late-phase peak to main
flare peak are still only 0.18 and 0.1 for the two cases in our experiments, because the long length of the late-phase loop significantly 
reduces the line peak irradiance from the loop. In diagnosing the flares from AR 11283, we also note that not all flares from this AR exhibit
an evident EUV late phase, even though the flare class is sufficiently high. In a real solar flare, it is very rare that the late-phase loops, if
they do exist, receive a total energy input comparable to or greater than that the input into the main flaring loops. Even if an EUV late 
phase is produced in the late-phase loops, the late-phase emission may be too weak to be observed. In addition to the reason pinpointed in 
\citet{LiYD14}, this may be another reason why EUV late-phase flares only occupy a small fraction in all solar flares. In passing, we note 
that \citet{LiuK15} have recently reported an extremely large EUV late phase (an even greater late-phase peak than the main flare 
peak) in a confined solar flare, attributing the production of this extremely large late phase to the persistent heating powered by a trapped 
hot structure (most presumably a flux rope).

There are still some other factors that we did not includ in our numerical modeling, which may potentially affect the loop emissions. 
The first one is nonthermal electron beam heating, whose effect has been addressed in \citet{LiuWJ13} by using the EBTEL model.
Nevertheless, many observations with \emph{RHESSI} \citep{LinRP02} have shown that HXR emissions, which 
are believed to be produced by the bremsstrahlung of nonthermal electrons in the solar lower atmosphere \citep[e.g.,][]{HaoQ17}, 
mainly come from the footpoints of short main flaring loops rather than long late-phase loops \citep[e.g.,][]{Feng13,SunX13}. 
The second factor is deviations from ionization equilibrium \citep{Reale08,Bradshaw11}. During the impulsive heating and conductive 
cooling stage, the temperature evolves so fast that the ionization process cannot catch up with the temperature variation, resulting in a 
non-equilibrium ionization. Obviously, this effect is the most prominent for hot coronal lines in long tenuous loops. As \citet{Bradshaw11} 
have pointed out, as the cooling rate slows down and the density increases, the ions can have enough time to reach an ionization
equilibrium at a medium temperature like 6~MK. Therefore the emissions formed below this temperature are unlikely to be notably 
affected by this effect. The last factor is CME-associated coronal dimming \citep{Aschwanden09,ChengJ16}. Of the two synthetic EUV 
late-phase flares, the late phase of the cool \ion{Fe}{9}  line in Case 1 is quite weak because of the fast mass draining in this stage, 
while the late phase in Case 2 is rather prominent. According to previous case studies \citep{Hock12,Dai13}, Case 2 is more likely to correspond
to an eruptive flare, in which the late-phase heating is caused by magnetic reconnection between the  magnetic field lines stretched out by 
the eruption of a flux rope near the time of main flare peak \citep{ChengX17}.  Mass depletion accompanying the CME lift-off will cause
a prolonged coronal dimming in the bulk coronal emission, which easily submerges the cool coronal late phase. Emissions of higher
temperatures, however, are little affected by this dimming, because they contribute only very little to the bulk coronal emission. 
To summarize, none of these factors may impose a notable effect on the EUV late-phase emission, as it comes from loops of great lengths
and medium temperatures. This may answer the question why the EUV late phase is mainly observed in warm coronal emissions.

\section{SUMMARY}
Using the EBTEL model, we have numerically synthesized two flare cases that mimic EUV late-phase solar flares produced via two 
main mechanisms, i.e., long-lasting cooling and secondary heating mechanisms. We probed in detail the physical link between the emission 
characteristics and the underlying hydrodynamics and thermodynamics in the flare loops. Our main conclusion is that the underlying 
hydrodynamic and thermodynamic evolutions in late-phase loops are different if they are generated by the two different mechanisms. 
The late-phase peak due to a long-lasting cooling process always occurs during the radiative cooling phase, while that powered by a 
secondary heating is more likely to take place in the conductive cooling phase. We then proposed a new method for diagnosing the two 
mechanisms based on the shape of late-phase light curves. The preliminary application of the method to real solar observations is encouraging. 
Moreover, we discussed the energy partition between the different loops in a solar flare, and pointed out that it is not easy for the flare 
to exhibit an evident EUV late phase. Finally, we also addressed some other factors that may potentially affect the loop emissions. We 
proposed that none of them may impose a notable effect on the warm coronal late-phase emission. 

To better understand the nature of the EUV late phase of solar flares, we need more observations. The MEGS-A component of EVE, 
whose spectral window covers the EUV lines used in this study, has been lost since 2014 May 26. Fortunately, we still have AIA in good 
working order. Because of the close similarity between the EUV emissions from the two instruments, as shown in this work, we can reliably
extract the late-phase information from a solar flare. 

\acknowledgements{We are very grateful of the anonymous referee for many valuable comments and suggestions. This work was 
supported by National Natural Science Foundation of China under grants 11533005 and  
11733003, and 973 Project of China under grant 2014CB744203. D.Y. is also sponsored by the Open Research Project
of National Center for Space Weather, China Meteorological Administration. The \emph{SDO} is a mission of NASA's Living With a
Star (LWS) Program.}


\end{document}